\documentclass[%
 reprint,
 amsmath,amssymb,
 aps,
]{revtex4-2}
\usepackage{lipsum}
\usepackage{graphicx}
\usepackage{dcolumn}
\usepackage{bm}

\begin{document}

\preprint{APS/123-QED}

\title{Transfer of Fisher Information in Quantum Postselection Metrology}

\author{Zi-Rui Zhong}
\author{Xia-Lin Su}
\author{Xiang-Ming Hu}
\author{Qing-Lin Wu}%
\email{qlwu@ccnu.edu.cn}
\affiliation{%
\emph{Department of Physics, Central China Normal University, Wuhan 430079, China}
}%
\affiliation{%
\emph{Key Laboratory of Quark \& Lepton Physics (MOE) and Institute of Particle Physics, Central China Normal University, Wuhan 430079, China}
}%

\date{\today}

\begin{abstract}
Postselected weak measurement has shown significant potential for detecting small physical effects due to its unique weak-value-amplification phenomenon. Previous works suggest that Heisenberg-limit precision can be attained using only the optical coherent states.  
However, the measurement object is the distribution of postselection, limiting the practical applicability. Here, we demonstrate that the output photons can also reach the quantum scale by utilizing the Fisher information transfer effect. In addition, we consider the insertion of a power-recycling cavity and demonstrate its positive impact on the distribution of postselection. Our results enhance the quantum metrological advantages of the postselection strategy and broaden its application scope.

\end{abstract}

\maketitle


\section{Introduction}
Postselected weak measurement, first introduced by Aharonov, Albert, and Vaidman (AAV) in 1988 \cite{PhysRevLett.60.1351}, has demonstrated rich potential in quantum mechanics, particularly in the area of parameter estimation and quantum measurement theory. It involves weak coupling between the system and the meter, enabling the average shift of output pointer state to significantly exceed the eigenvalues of the system observable throughout postselection \cite{steinberg2010}. Generally, the amplification of the shift observable is associated with the concept of the 'weak value,' a phenomenon known as the weak-value-amplification (WVA) effect. 
With these characteristics, postselected weak measurement has been successfully used in the detection of tiny physical effects \cite{science.1152697,PhysRevA.85.043809,PhysRevLett.109.013901,PhysRevLett.102.173601,PhysRevA.80.041803,PhysRevA.82.063822,PhysRevLett.111.033604,Viza:13,Santana:16,Egan:12,10.1063/1.5027117,PhysRevLett.112.200401,DELIMABERNARDO20142029,Fang:21,Qiu2017,Bai:20,Qu2020,PhysRevLett.102.020404}. 

The accuracy of a measurement is a crucial factor in assessing the effectiveness of a measurement method. 
Whether postselected weak measurement is optimal remains a significant debate \cite{PhysRevA.85.062108,PhysRevA.85.060102,PhysRevX.4.011031,PhysRevA.88.042116,PhysRevLett.107.133603,PhysRevLett.114.210801,Jordan2015,PhysRevLett.112.040406,PhysRevA.106.022619,PhysRevA.102.042601,PhysRevA.91.032116,PhysRevX.4.011032,PhysRevA.91.062107,PhysRevA.96.052128}. Critics often address this issue from the perspective of Fisher information, arguing that postselection, in itself, cannot improve accuracy and instead discards a significant portion of the photons that carry useful information\cite{PhysRevLett.112.040406}. Positive discussions focus on specific conditions. For example, weak measurement can outperform the conventional measurement in the presence of detector saturation \cite{PhysRevLett.118.070802,PhysRevLett.125.080501}. Additionally, weak measurement has been pointed out to suppress the technique noise in some circumstances \cite{PhysRevLett.105.010405,PhysRevA.85.060102,PhysRevA.102.042601,PhysRevX.4.011031,RevModPhys.86.307,PhysRevLett.132.043601,PhysRevA.92.012120}. Another interesting theoretical consideration is that postselected experiments can yield anomalously large information-cost rates \cite{Arvidsson-Shukur2020}. 

Since debates on weak measurement often center around low postselection efficiency, several targeted efforts have been made to address this issue. Recycling techniques have been demonstrated to enhance both the postselection probability and signal-to-noise ratio by reusing the failed postselection photons \cite{PhysRevLett.114.170801,PhysRevA.88.023821,PhysRevLett.117.230801,FANG2020125117,PhysRevLett.126.220801,PhysRevA.108.032608,PhysRevA.109.042602}. Additionally, joint weak measurement has been shown to maximize photon utilization and remain robust against various sources of noise \cite{PhysRevLett.110.083605}. Furthermore, certain quantum resources, such as entanglement and squeezing, can enhance the precision of WWA measurements to the quantum scale \cite{PhysRevLett.113.030401, PhysRevA.92.012120, PhysRevLett.115.120401, PhysRevA.99.032120, PhysRevLett.115.120401}. Some schemes even claim to achieve Heisenberg-scaling precision without relying on entanglement \cite{PhysRevLett.128.040503} or any quantum resources \cite{PhysRevLett.114.210801, Jordan2015, PhysRevLett.121.060506,Chen2018}. These works undoubtedly provide a broader perspective on the quantum potential of WWA-based metrology. 

In this article, we focus on the WWA setup presented in Ref. \cite{PhysRevLett.114.210801}, where the distribution of postselection can yield quantum-enhanced precision. We will demonstrate that the successful or failed postselection states can also reach the quantum scale, utilizing the concentration or transfer effect of Fisher information in weak measurement. In addition, we will show recycling techniques can enable the postselection process itself to benefit, rather than just the successful postselection photons. 

\begin{figure}[t]
\centering
\includegraphics[trim= 0.1 0.1 0.1 0.1 ,clip, scale=0.6]{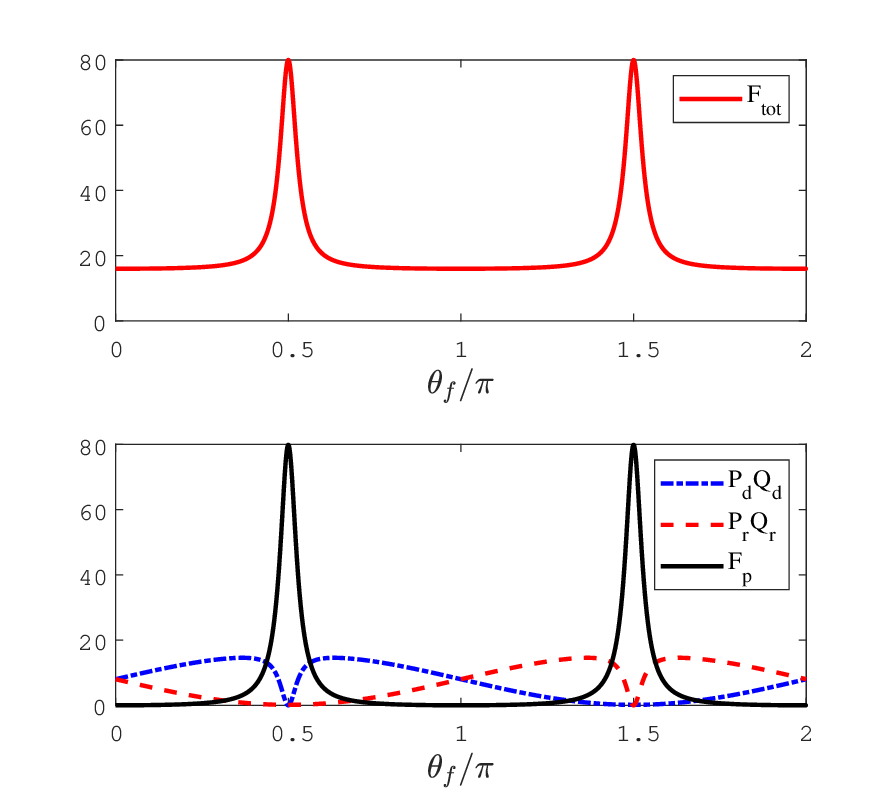}
\caption{(Color online) Fisher Information (FI) Composition. Panel (a) corresponds to the cases of the total FI $F_{tot}$ varying with $\theta_f$. Panel (b) corresponds to the cases of the separated FI $P_dQ_d$, $P_rQ_r$ and $F_p$ varying with $\theta_f$.   
In these plots, the average photon number is set to $\bar{n}=4$. $F_{tot} $is the sum total of the FI contained at different stages of the preselection-coupling-postselection process. $P_dQ_d$ is the quantum Fisher information (QFI) of successful postselection photons. $P_rQ_r$ is the QFI of failed postselection photons. $F_p$ is the FI of the distribution of postselection probability} 
\label{Fig.1}
\end{figure}

\begin{figure*}[t]
\centering
\includegraphics[trim= 0.1 0.1 0.1 0.1 ,clip, scale=0.7]{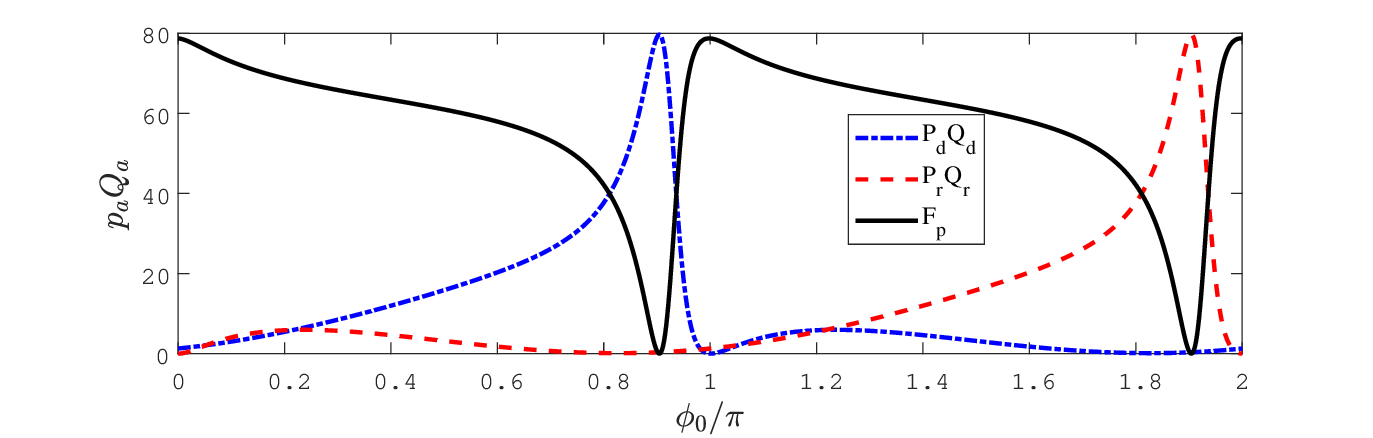}
\caption{(Color online) Transfer of Fisher Information (FI). $P_dQ_d$, $P_rQ_r$ and $F_p$ vary with the phase shift between the pre- and post-selected state $\phi_0$. In this plot, all $P_dQ_d$, $P_rQ_r$ and $F_p$ can achieve the maximum value of $4\bar{n}^2+4\bar{n}$. $F_{tot} $is the sum total of the FI contained at different stages of the preselection-coupling-postselection process. $P_dQ_d$ is the quantum Fisher information (QFI) of successful postselection photons. $P_rQ_r$ is the QFI of failed postselection photons. $F_p$ is the FI of the distribution of postselection probability}
\label{fig.2}
\end{figure*}

\section{Fisher information analysis on postselection measurement using pure coherent states}
In this section, we review the WWA-based setup introduced by Zhang \textit{et. al} \cite{PhysRevLett.114.210801}. The quantum system is a two-state system with eigenstates $|e\rangle$, $|g\rangle$, which is coupled to an optical meter prepared in a coherent state $|\alpha\rangle$. The initial state of quantum system is prepared as $|\psi_{i}\rangle=\cos{(\theta_i/2)}|g\rangle+\sin{(\theta_i/2)}e^{i\phi_{i}}|e\rangle$. The interaction Hamiltonian produces a unitary operation represented as follows: 
\begin{equation}
    U=\exp{(ig\hat{\sigma}_{z}\hat{n})},
\end{equation}
where $g$ is the coupling parameter, $\hat{\sigma}_{z}$ is the Pauli operator $\hat{\sigma}_{z}=|e\rangle \langle e| - |g\rangle \langle g|$, and $\hat{n}$ is the photon-number operator $\hat{n}=\hat{a}^{\dagger}\hat{a}$. This results in the joint state to be
\begin{equation}
    \begin{aligned}|
    \Psi_{j}\rangle=U|\psi_{i}\rangle |\alpha\rangle=\cos\frac{\theta_{i}}{2}|g\rangle|\alpha e^{-ig}\rangle+\sin\frac{\theta_{i}}{2}e^{i\phi_{i}}|e\rangle|\alpha e^{ig}\rangle.
    \end{aligned}
\end{equation}
Postseleting the system in the state $|\psi_{f}\rangle=\cos{(\theta_f/2)}|g\rangle+\sin{(\theta_f/2)}e^{i\phi_{f}}|e\rangle $, the  meter state becomes
\begin{equation}
|\widetilde{\Phi}_d\rangle=\cos\frac{\theta_{i}}{2}\cos\frac{\theta_{f}}{2}|\alpha e^{-ig}\rangle+\sin\frac{\theta_{i}}{2}\sin\frac{\theta_{f}}{2}e^{i\phi_{0}}|\alpha e^{ig}\rangle, 
\end{equation}
where $\phi_0=\phi_i-\phi_f$ denotes the phase shift between the initial system state and postselected system state. 
Notably, $|\widetilde{\Phi}_d\rangle$ is not normalized. The related normalized form should be $|\Phi_d\rangle=|\widetilde{\Phi}_d\rangle/\sqrt{P_{d}}$, where the normalization factors are given by $P_d=\langle\widetilde{\Phi}_{d}|\widetilde{\Phi}_{d}\rangle$, which is also the probability of successful postselection. 
Similarly, the failed postselected state can be written as $|\psi_{r}\rangle=\sin{(\theta_f/2)}|g\rangle-\cos{(\theta_f/2)}e^{i\phi_f}|e\rangle $, corresponding to the reflected probe, 
\begin{equation}
|\widetilde{\Phi}_r\rangle=\cos\frac{\theta_{i}}{2}\cos\frac{\theta_{f}}{2}|\alpha e^{-ig}\rangle-\sin\frac{\theta_{i}}{2}\sin\frac{\theta_{f}}{2}e^{i\phi_{0}}|\alpha e^{ig}\rangle. 
\end{equation}
Thus, the corresponding normalized forms is $|\Phi_r\rangle=|\widetilde{\Phi}_r\rangle/\sqrt{P_{r}}$, where $P_r=\langle\widetilde{\Phi}_{r}|\widetilde{\Phi}_{r}\rangle$ is the probability of failed postselection. 

The Fisher information (FI) is a central quantity in parameter estimation, denoted by $F$ in this article. 
It establishes a lower bound on the variance of an unbiased estimator $g$ through the Cramér-Rao inequality: $Var(g)\ge 1/NF$\cite{Wiseman_Milburn_2009}.  For a given probability distribution $p_m$, we can calculate the FI via the formula 
\begin{equation}\label{e1}
    F=\sum_{m}\frac{1}{p_m}\left(\frac{dp_m}{dg}\right)^2. 
\end{equation}
The quantum Fisher information (QFI) is defined as the FI maximized over all possible generalized measurements. When the meter function $|\Phi\rangle$ is known, the QFI can be calculated using the following formula
\begin{equation}\label{e2}
   Q= 4\left[(\frac{d\left \langle \Phi\right | }{dg } )(\frac{d\left | \Phi  \right \rangle }{dg } ) - {\left |\left \langle \Phi \right| (\frac{d\left | \Phi  \right \rangle }{dg } ) \right |}^2 \right].
\end{equation}
By substituting $|\Phi_d\rangle$ and $|\Phi_r\rangle$ into Eq. (\ref{e2}), we obtain the QFI of successful and failed postselection parts, denoted as $Q_d$ and $Q_r$, respectively. $P_dQ_d+P_rQ_r$ stands for the maximum information from detecting photons. Besides, the distribution $\{P_d,P_r\}$ yeilds a classical FI $F_p$, which refers to the information obtained in the postselection process. Therefore, the total information contained  at different stages of the preseletion-coupling-postseletion process is 
\begin{equation}
    F_{tot}=P_dQ_d+P_rQ_r+F_p, 
\end{equation}
which is also one of the major conclusions of \cite{PhysRevLett.114.210801}. 

Next, we will demonstrate how to achieve the maximum $F_{tot}$. Notably, $F_{tot}$ cannot exceed the input Fisher information $Q_j$, which can be obtained by substituting $|\Psi_j\rangle$ into Eq. (\ref{e2}). It can be calculated that $Q_j=4n^2\sin^2 {\theta_i}+4n$, where $n=|\alpha|^2$ is the average photon number. 
This expression suggests that $\theta_i=\pi/2$ can provide the maximum value $Q_j=4n^2+4n$, where the upper limit of $F_{tot}$ also reaches its maximum.
Next, we examine the optimal region of $F_{tot}$. As shown in Fig. \ref{Fig.1}, we plot $F_{tot}$, $P_dQ_d$, $P_rQ_r$ and $F_p$ varying with $\theta_f$ where we set $n=4$ and $\phi_0=\pi$. We can clearly see that the peak $F_{tot}$ corresponds to $\theta_f=\pi/2$ or $3\pi/2$ (considering only within one cycle), and the Fisher information is concentrated in the postselection distribution. In this way, both $Q_j$ and maximum $F_{tot}$ can exhibit quantum scaling ($\sim n^2$). 

However, the quantum scale only occurs in the classical FI of the postselected process under the parameter conditions specified in previous literature \cite{PhysRevLett.114.210801,Jordan2015,PhysRevLett.121.060506,PhysRevA.106.022619}. In other words, we need to establish a measurement scheme based on the postselection distribution rather than straightly detecting the output state. This limitation affects the practical applicability of this approach. So here raises a immediate question: can the output meter state ($P_dQ_d$) reaches quantum scale? 

\section{The information transfer effect of weak-value measurement}

To begin with, we must reiterate the concept of weak value \cite{PhysRevLett.60.1351}, which is typically employed to amplify the measured parameters. Its mathematical form here is 
\begin{equation}
    A_w = \frac{\langle \psi_f | \hat{\sigma}_z  | \psi_i   \rangle}{\langle \psi_f  | \psi_i   \rangle}
     = \frac{\sin{\theta_i}\sin{\theta_f}e^{i\phi_0}-\cos{\theta_i}\cos{\theta_f}}{\sin{\theta_i}\sin{\theta_f}e^{i\phi_0}+\cos{\theta_i}\cos{\theta_f}}.
\end{equation}
Pervious related works conclude that weak-value measurement can concentrate the input FI regarding the measured parameter into a small successful-postselection events \cite{PhysRevX.4.011031,PhysRevLett.113.030401,PhysRevLett.114.170801}. 
However, in Ref. \cite{PhysRevLett.114.210801}, the FI is concentrated in the postselection distribution rather than in the typical aggregation of successful-postselection photons. 
This change stems from an alteration in the parameter region. For example, by focusing on one of the maximum $F_{tot}$ parameter conditions where $\theta_i = \theta_f = \pi/2$, we can readily calculate that $A_w = 1$. However, this does not represent a strict weak-value measurement, as there is no manifestation of the weak-value amplification effect. Reference \cite{Jordan2015} provides a more in-depth explanation of this issue; here, we merely emphasize that this measurement method is distinct from traditional weak-value measurement. 

Next, we will demonstrate how this distinction significantly influences the postselection. 
we set $\theta_i = \theta_f = \pi/2$, which corresponds to the parameter condition for the maximum $F_{tot}$ (this condition will be consistently applied in the following sections), and present the equation for the probability of successful postselection
\begin{equation}\label{eq9}
    P_d=\frac{1+e^{-2n\sin^2{g}}\cos{(n\sin{2g}+\phi_0)}}{2}. 
\end{equation}
When we apply the parameter settings from Reference \cite{PhysRevLett.114.210801}, where $\phi_0 = \pi$ and a constant incident photon number, the post-selection is exclusively introduced solely by the weak coupling $g$. However, for typical weak-value measurements, where the angle between the pre- and post-selected states is $\phi$ and the the parameter range is $g \ll \phi \ll 1$, the post-selection process is primarily determined by $\phi$, rather than by $g$. In other words, as $\phi$ increases from $0$ into the weak-value region, the FI is continuously transferred from the postselection process to the successful-postselection photons. 

Finally, we will demonstrate this transfer effect also exists in our scheme. From Eq. (\ref{eq9}), disregarding high order terms of $g$, it is evident that $\phi_0$ holds the same significance as $n\sin{2g}$. To ensure how $\phi_0$ affects the postselection, we plot  $P_dQ_d$, $P_rQ_r$ and $F_p$ varying with $\phi_0$ in Fig. \ref{fig.2}, where the trend of FI transfer is clearly illustrated. 
It can be seen that $P_dQ_d$ can reach the maximum scale $4n^2+4n$, with the peak corresponding to the horizontal coordinate $\phi_0=\pi-n\sin{2g}-2g$. (The calculation here is straightforward, as $F_p$ decreases to $0$ at the peak value of $P_dQ_d$. We simply need to solve for $F_p=0$.) 
Therefore, our conclusion is that during the transition of $\phi_0$ from $\phi_0=\pi$ to $\phi_0=\pi-n\sin{2g}-2g$, the Fisher information gradually transfers from the postselection distribution to the successful-postselection photons, resulting in the output photons $\sim n^2$ scaling. 

\begin{figure}[t]
\centering
\includegraphics[trim= 2 2 2 2 ,clip, scale=0.45]{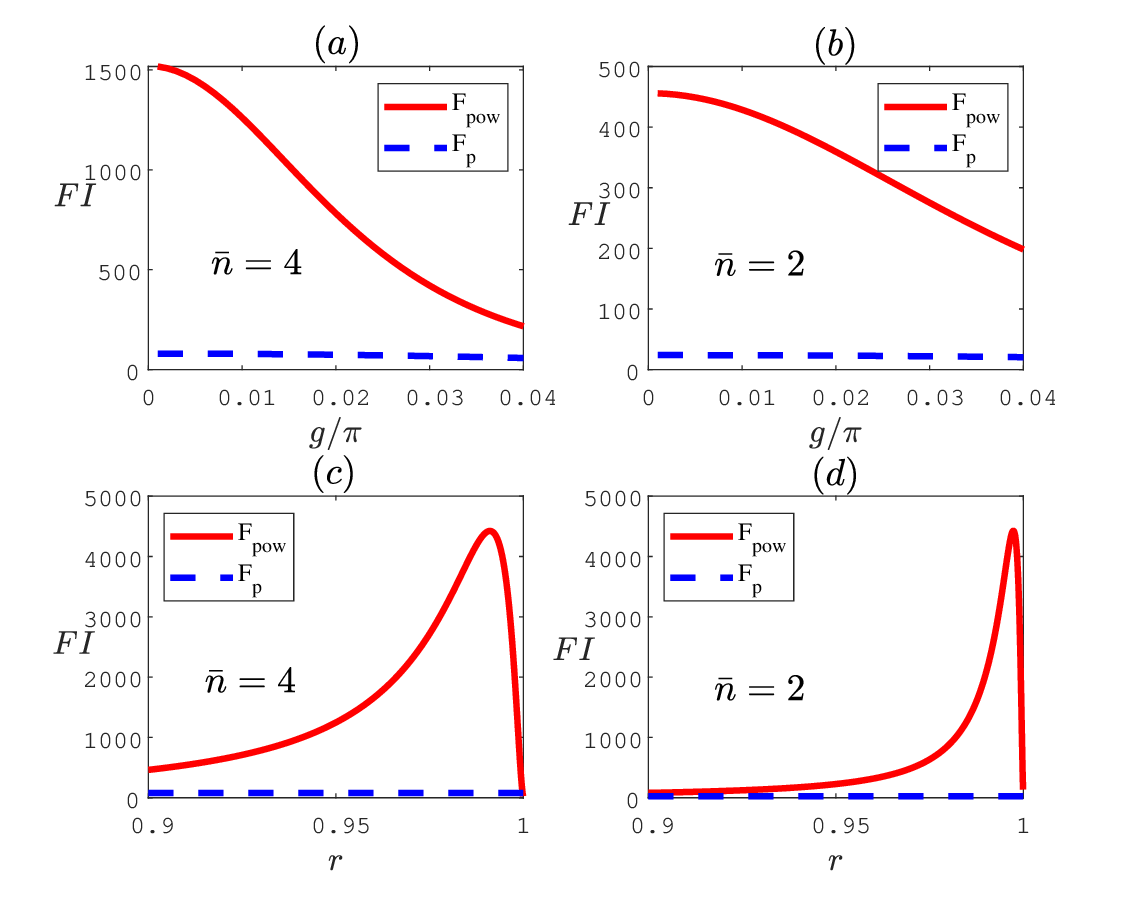}
\caption{(Color online) Comparison of postselection in power-recycling and standard WWA schemes. (a) and (b) correspond to the case of $F_{pow}$ and $F_p$ varying with $g$ under $\bar{n}=4,\ 2$, respectively, where $r=0.9$. In (c) and (d), we assume $g=0.01\pi$ and plot $F_{pow}$ and $F_p$ varying with $r$ under $\bar{n}=4,\ 2$, respectively. $F_{pow}$ and $F_p$ are the Fisher information of the postselection process of power-recycling and standard schemes, respectively. $g$ is the coupling parameter. $r$ is the reflection coefficient of mirror. }
\label{fig.3}
\end{figure}

\begin{figure}[t]
\centering
\includegraphics[trim= 2 2 2 2 ,clip, scale=0.6]{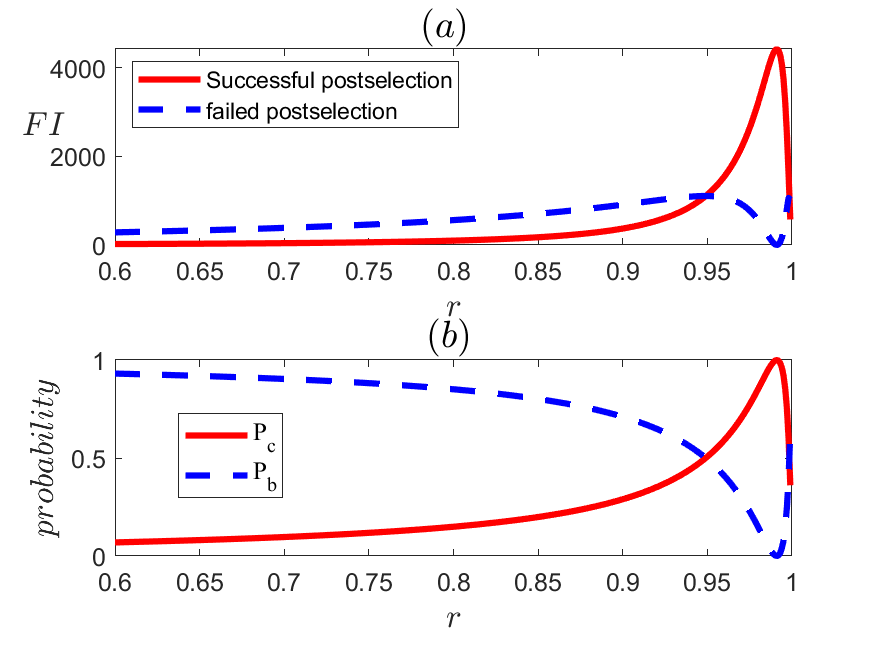}
\caption{(Color online) The origin of the benefits of postselection induced by power-recycling. 
(a) The Fisher information of successful and failed postselection parts varying with $r$ under $\bar{n}=4$ and $g=0.01*\pi$. (b) The trend of successful and failed postselection probabilities. 
$F_{pow}$ and $F_p$ are the Fisher information of the postselection process of power-recycling and standard schemes, respectively. $g$ is the coupling parameter. $r$ is the reflection coefficient of mirror. }
\label{fig.4}
\end{figure}

\section{power-recycling can benefit the distribution of postselection}

The power-recycling method effectively utilizes the information concentration effect of weak-value measurement. It transforms failed-postselection photons, which carry low information, into successful-postselection light with high information through a resonant cavity. 
Theoretically, it can enhance the Fisher information of the outgoing photons by a factor of $1/\chi$, where $\chi$ is proportional to $\sqrt{P_d}$ \cite{PhysRevX.4.011031,PhysRevLett.114.170801}. There is no doubt that this enhancement also applies to the coherent meter state here. 
However, during this photon transfer process, the postselection distribution undergoes a change.
We will demonstrate that this change is also beneficial. 

Power-recycling technology involves inserting a partially transmitting mirror with reflection and transmission coefficients respectively $r$ and $p$ ($r^2+p^2=1$).
Meanwhile, the post-selection can be regarded as a specialized partially transmitting mirror, since it allows a small portion of photons to reach the detector, while the majority are reflected back to the laser. 
Here, we assume the reflection coefficient of \textit{postselection} are $r_m$, which satisfies $r_m=\langle \psi_i|U| \psi_i \rangle = \sqrt{P_r} $. So the corresponding transmission coefficient $p_m$ can be given by $p_m=\sqrt{1-r_m}$. 
Considering an initial amplitude $E_0$ incident into the system, the amplitudes of reflected and transmitted light become $E_{rm}=r_mE_0$ and $E_{rt}=p_mE_0$, respectively, where we neglect the optical loss. 
Therefore, the light inside the cavity and reflected back to the laser are respectively 
\begin{equation}  \label{ea1}
    \begin{aligned}
    E_{c}&=p\left [ 1 + rr_me^{i\theta} + \left(  rr_me^{i\theta}\right)^2 
    + \dots\right]E_0\\
    &=\frac{p}{1-rr_me^{i\theta}}E_0
    \end{aligned}
\end{equation}
and
\begin{equation}
    \begin{aligned}
    E_{b}=\left( -r+\frac{p^2r_me^{i\theta}}{1-rr_me^{i\theta}} \right)E_0, 
    \end{aligned}
\end{equation}
where $\theta$ is the single-pass phase addition. 
For resonance cavity, $e^{i\theta} \approx 1$, and the gain factor of amplitude is $E_{cav}/E_0=p/(1-rr_m)$. This leads to the successful-postselection probability changing from $p_d$ to $p_c$, 
\begin{equation}
    P_c=\left( \frac{p}{1-r\sqrt{P_r}}\right)^2P_d. 
\end{equation}
Similarly, the probability of failed postselection photons can be expressed as the square of the unnormalized amplitude of the light reflected back toward the laser.
\begin{equation}
    P_b=\left( -r+\frac{p^2\sqrt{P_r}}{1-r\sqrt{P_r}} \right)^2. 
\end{equation}
By substituting the distribution of $\{P_c,P_b\}$ into Eq. (\ref{e1}), we can obtain the FI of postselection process modified by power-recycling 
\begin{equation}
    F_{pow}=\frac{1}{P_c}\left(\frac{dP_c}{dg}\right)^2+\frac{1}{P_b}\left(\frac{dP_b}{dg}\right)^2. 
\end{equation}

To determine whether power-recycling affects the distribution of postselection, we first choose the optimal parameter  $\phi=\pi$, which corresponds to the maximum $F_p$. Additionally, we assume $\sin{g}\approx g$ and neglect high-order terms of $g$ ($g \ll 1$), thus simplifying $F_{pow}$ to 
\begin{equation}\label{eq14}
    F_{pow}\approx n^2B^2\left[ 1 + \frac{\left(\cos{ng}-r\right)^2}{\left(1-r^2\right)\sin^2{ng}} \right], 
\end{equation}
where
\begin{equation}
    B=\frac{(1-r^2)(\sin{2ng}+2g\cos{2ng})}{(r\cos{ng}-1)^2\cos{ng}}. 
\end{equation}
It is evident that the power-recycled distribution of postselection can also achieve quantum precision ($\sim n^2$). To visualize this, 
as shown in Fig. \ref{fig.3}, we plot the curves of $F_{pow}$ (red solid line) and $F_p$ (blue dashed line), where $F_{pow}$ exhibits a broad-range enhancement. This implies that probability distribution $\{P_c,P_b\}$ contains significantly more FI than probability distribution $\{P_d,P_r\}$. 

In addition, to analyze the source of the improvement in Fisher information, we divide $F_{tot}$ into two parts, 
\begin{equation}
    F_{tot}=F_c+F_b, 
\end{equation}
where $F_c$ is the contribution from the successful postselection part and $F_b$ is the contribution from the failed postselection part. 
As shown in Fig. \ref{fig.4}, we compare the trends of change in Fisher information and postselection probabilities. We obtained two similar trends, but the magnitudes of these trends differ significantly. The conclusions are as follows:
(i) The successful postselection part has a larger information-to-probability ratio than failed postselection part;
(ii) Power-recycling cavity transfers photons from the failed postselection part to the successful postselection part, thereby increasing the Fisher information of postselection.

\section{Conclusion}
In summary, we have considered the postselected metrology model proposed by Zhang \textit{et al}. \cite{PhysRevLett.114.210801} and have analyzed the effect of Fisher information transfer within it. We have found that the Heisenberg-scale Fisher information can be transferred from the distribution of postselection probability to the output meter state. Furthermore, we have introduced a power-recycling cavity and have found that it significantly optimizes the distribution of postselection by transferring photons from the low-efficiency part to the high-efficiency part. 

This Fisher information transfer effect originates from posselection, and therefore, it theoretically manifests in all postselection-related measurements. Further work may focus on combining this ideal model with various types of noise, such as quantum dephasing. 

\section{acknowledgements}

This work was financially supported by self-determined research funds of CCNU from the colleges' basic research and operation of MOE and the National Natural Science Foundation of China (Grants No. 61875067).
\nocite{*}

\appendix

\end{document}